\documentclass[11pt, oneside]{article}

\usepackage[a4paper, total={6.2in, 9in}]{geometry}
\usepackage[utf8]{inputenc}
\pdfoutput=1
\usepackage{graphicx}
\usepackage{booktabs}
\usepackage{ragged2e}
\usepackage{multirow}
\usepackage{multicol}
\usepackage{pbox}
\usepackage[utf8]{inputenc}
\usepackage{array}
\usepackage{tabularx}
\usepackage{caption}
\usepackage{rotating}
\usepackage{amssymb}
\usepackage{xcolor}
\usepackage[T1]{fontenc}
\usepackage{lmodern}
\usepackage[bottom, multiple]{footmisc}
\usepackage[normalem]{ulem} % for \sout (line-throughs)

\PassOptionsToPackage{hyphens}{url}\usepackage[hidelinks]{hyperref}
\usepackage{fancyhdr}                           % Needed for footers on each page
\fancypagestyle{firstpage} % first page footer only
{
   \fancyhf{}
   \fancyfoot[C]{\vspace{-12mm}\thepage}
   \fancyfoot[L]{\vspace{0mm}\noindent\rule{6.2in}{0.4pt}
     \scriptsize{
       *This paper appeared in the Journal of Financial Market Infrastructures Volume 9, Number 4 (June 2021), pages 51–62. \newline
         \textcopyright \hspace{0.5mm} Barclays Bank PLC 2020 \\
       This work is licensed under a \href{https://creativecommons.org/licenses/by/4.0/}{Creative Commons Attribution 4.0 International License (CC
       BY).}  \\ Provided you adhere to the CC BY license, including as to attribution, you are
       free to copy and redistribute this work in any medium or format and remix, transform,
       and build upon the work for any purpose, even commercially.                  \\
       BARCLAYS is a registered trademark of Barclays Bank PLC, all rights are reserved. \\
     }
 }
}
\title{\vspace{-2.5cm}Industry Adoption Scenarios for Authoritative Data Stores \\ using the International Swaps and Derivatives Association Common Domain Model*}

\usepackage{anyfontsize}

\begin{document}
\author{
  \hspace{-1.5em}
  \begin{tabular}{c} {\fontsize{10.75}{1cm}\selectfont \hspace{3mm} Aishwarya Nair and Lee Braine} \\ {\fontsize{10.75}{1cm}\selectfont \hspace{4mm} Chief Technology Office} \\ 
  {\fontsize{10.75}{1cm}\selectfont \hspace{4mm} Barclays} \end{tabular}
}

\date{July 13, 2020\\ {\fontsize{8}{1cm}\selectfont (Revised August 9, 2022)}}

% Getting rid of line breaks being makde with hyphens
\pretolerance=10000
\tolerance=2000 
\emergencystretch=10pt

%\begin{document}
\maketitle
\thispagestyle{firstpage} % Needed to get footer on first page
\vspace{-1cm}
\begin{abstract}
\noindent
In this paper we explore opportunities for the post-trade industry to standardize and
simplify in order to significantly increase efficiency and reduce costs. We start by
summarizing relevant industry problems (inconsistent processes, inconsistent data
and duplicated data) and then present the corresponding potential industry solutions
(process standardization, data standardization and authoritative data stores).
This includes transitioning to the International Swaps and Derivatives Association
Common Domain Model (CDM) as a standard set of digital representations for the
business events and processes throughout the life cycle of a trade. We then explore
how financial market infrastructures could operate authoritative data stores that make
CDM business events available to broker-dealers, considering both traditional centralized
models and potential decentralized models. For both types of model, there
are many possible adoption scenarios (depending on each broker-dealer’s degree of
integration with the authoritative data store and usage of the CDM), and we identify
some of the key scenarios.

\end{abstract}

\section{Introduction}
\label{sec:introduction}
\noindent
The stages of the trade life cycle include client onboarding, execution and post-trade.
This paper focuses on post-trade, which refers to the back- and middle-office activities
after a trade is executed. These activities include matching, confirmation, trade
enrichment, collateral and margin, trade settling and certain life-cycle events and
corporate actions. The post-trade industry is large (with an estimated US\$20 billion
total annual spending on post-trade processes \cite{boe-future-post-trade}) and complex
(partly due to the incremental evolution of post-trade processes and technology over
decades). There are significant opportunities to simplify and reduce costs.
Here, we aim to provide a design discussion of relevance to financial institutions,
trade associations, fintechs and official institutions. We consider some important
industry problems (inconsistent processes, inconsistent data and duplicated data),
which have become key obstacles to improving post-trade efficiency. We then identify
the corresponding potential industry solutions, including the use of the International
Swaps and Derivatives Association (ISDA) Common Domain Model (CDM) \cite{isda-cdm-factsheet} and an authoritative data
store (ADS). We also provide high-level descriptions with architecture diagrams to
illustrate some of the key adoption scenarios for financial institutions, with the intention
of conveying a sense of the variety of architectural options for both centralized
and decentralized models in this design space. We hope the topics raised in this paper
will stimulate discussion and we look forward to continuing industry collaboration
on ADSs and the CDM.

\pagebreak

\section{Industry Problems}
\label{sec:industryProblems}
\noindent
The industry’s current methods of managing trades through the post-trade life cycle
can be inefficient. For example, there are estimated potential cost savings of 80\% 
from just the dealer cost base of approximately US\$3.2 billion within the primary
areas directly impacted by the CDM \cite{deloitte-future-of-post-trade}. 
In addition, it has been noted that if the cost per trade and the cost of doing business become unacceptably
high, then financial institutions may start exploring the ongoing viability of certain
offerings \cite{isda-quaterly}. It is therefore of strategic importance that the fundamental problems are identified, acknowledged and
addressed by the industry.

We summarize the three fundamental industry problems as follows.

\begin{itemize}
\item {\em Inconsistent processes.} Over time, each financial institution has separately developed
a large number of complex post-trade business processes to support functions
that are essentially the same both within and across asset classes \cite{isda-future-of-derivatives}. The resulting variation across the industry
in processing business logic and life-cycle events has introduced significant
inefficiencies and increased the risk of errors requiring remediation. 

\item {\em Inconsistent data.} Over time, each financial institution has separately enriched trade
data and reference data with custom fields and values to incorporate the additional
information required by their custom processes. The resulting variation across the
industry in both data formats and data values has, similarly to the variation in
business processes, introduced significant inefficiencies and increased the risk of
errors requiring remediation.

\item {\em Duplicated data.} Trade data is stored in multiple entities across the industry, eg, at
each of the counterparties and at a central counterparty (CCP). In addition, partly
as a side effect of satisfying additional requirements over many years, changes
to existing infrastructure have sometimes included storing trade data in multiple
locations within each entity. Such data duplication across the industry clearly
introduces inefficiencies.
\end{itemize}

\noindent
Therefore, there is a perfect storm of industry inefficiency in post-trade processing,
fueled by duplicated, inconsistent processes operating on duplicated trade data in
inconsistent data formats. One result of this situation is that, every time there is a
life-cycle event, each of the copies of the trade may need to be updated and then
reconciled with each other \cite{isda-quaterly}. However, there are potential industry solutions to these industry problems and the
next section discusses a promising way forward.

\section{Potential Industry Solutions}
\label{sec:PotentialIndustrySolutions}
\noindent
The industry problems identified above (inconsistent processes, inconsistent data
and duplicated data) could be resolved via the rigorous adoption of process standards,
data standards and ADSs. Given the nature and scale of these industry problems,
the potential industry solution must be bold. Transformation across both financial
market infrastructures (FMIs)\footnote{We use the term FMI to refer to a “legal or functional entity that is set up to carry out centralised,
multilateral payment, clearing, settlement, or recording activities” and we “exclude [from
the definition] the participants that use the system” \cite{bis-principles-for-fmis}.} 
and broker-dealers\footnote{We use the term broker-dealer to refer to “any person engaged in the business of effecting transactions
in securities for the account of others” (a broker) or “any person engaged in the business of buying and selling securities 
for his own account, through a broker or otherwise” (a dealer) \cite{sec-guide-to-broker-dealer-registration}.}  covering processes, data formats and 
data repositories would clearly be a significant and lengthy endeavor. The
adoption challenges would include industry coordination on business cases, funding/
resourcing, prioritization, deployment and potentially the decommissioning of legacy
systems. Therefore, in this paper we explore the incremental stepping-stones on the
route toward a desirable end state.

\subsection{Standardisation}
\label{sec:Standardisation}
\noindent
The current lack of standardization could be ameliorated via industry-wide adoption
of the ISDA CDM. The industry already has a standard file exchange format called
the Financial products Markup Language (FpML) \cite{fpml-standard}, but that does not address the
problem of process variation resulting from custom business logic and custom calculations.
Building on the principles of FpML, the CDM in addition provides process
standardization.

ISDA published an initial digital representation of the CDM in 2018, providing
“a standard digital representation of events and actions that occur during the life of a
derivatives trade” \cite{isda-publishes-cdm-2018}. In 2019,
ISDA and REGnosys provided open access to the “full version of the CDM for interest
rate and credit derivatives” \cite{REGnosys}. Since then, the model has been
downloaded, deployed and tested by many market participants. For example, Barclays
(in collaboration with ISDA and REGnosys) hosted the Barclays DerivHack
2018 coding hackathon for market participants to explore the CDM for derivatives \cite{deloitte-future-of-post-trade} 
as well as the Barclays DerivHack 2019 for securities \cite{ey-fin-services-can-accelerate}. There are several industry working groups with supportive market participants
(including financial institutions, fintechs and law firms) contributing to the CDM; for
example, ISDA has CDM working groups for collateral and equity \cite{ISDA-Committees}, and the ICMA
has a CDM working group for bonds and repo markets \cite{icma-cdm}. However, broad industry
adoption will require an increasing number of market participants to engage in the
coming years.

Further, ISDA is working to directly integrate more of its legal documentation,
such as the Credit Support Documents, the new 2020 Interest Rate Definitions \cite{isda-from-2006-to-2020, isda-impetus-for-automation} 
and the ISDA Clause Library \cite{isda-clause-library}, with the CDM via legal-agreement components in the
model. This will allow the consistent implementation of operational clauses and
related procedures from ISDA documentation, thus tying legal data to systems that perform activities governed by legal contracts, permitting analyses that can improve
legal risk management.

\subsection{ADSs}
\label{sec:ADSs}
\noindent
The industry problem of duplicated data could be ameliorated via industry-wide
adoption of ADSs. An ADS can be considered a primary source of information that
acts as a single logical reference point, thereby avoiding the inconsistencies that
can arise in duplicated data. An ADS can be implemented using either a centralized
model or a decentralized model, depending on requirements such as resilience, data
distribution and technology maturity. Significant effort and resources are required to
build an ADS for an individual financial institution and so, when considering an ADS
for the industry, broker-dealers may look to FMIs to help drive industry adoption over
several years.

\section{Architecture Models}
\label{sec:ArchitectureModel}
\noindent
Building on the potential industry solutions discussed above, we now explore how
ADSs could be deployed to make CDM business events available to broker-dealers.
For simplicity, we assume each ADS would be operated by an FMI. We take a high-level
architecture perspective and consider both a traditional centralized model and
a potential decentralized model. For each model, there are many possible adoption
scenarios (depending on each broker-dealer’s degree of integration with the ADS
and usage of the CDM) and we explore four key adoption scenarios as architecture
options. In both models, the first three scenarios can be considered interim stepping-stones
and the fourth scenario can be considered an end state. A particular broker-dealer
may decide to remain on an interim scenario (for operational or other reasons),
although they would not benefit from the simplifications and efficiencies of the end
state. Our working assumption is that such architecture options could, if required, all
coexist within a particular infrastructure system, thereby permitting broker-dealers
with different degrees of integration to all participate in the same system.

\subsection{Centralized Model}
\label{sec:centralizedModel}
\noindent
In a centralized model, the FMI maintains a central ADS, as shown in Figure 1. The
FMI receives trade submissions (eg, from electronic communication networks and
via voice trading), processes the trades internally and carries through CDM business
events to the central ADS. Broker-dealers can then consume the CDM business
events via an application programming interface (API). Each broker-dealer maintains
a local copy of relevant data from the central ADS, implemented in a form that
could potentially range (depending on requirements) from a simple log to a database containing 
replicated trades. The broker-dealer also maintains relevant local controls
and services (including trade process data that is private to the broker-dealer). The
combination of the local copy of the ADS and the local controls and services can
be considered the broker-dealer’s ADS. There are several implementation options to
create this combination, such as enriching private fields in CDM business events with
broker-dealer-specific data or joining “immutable” trades with broker-dealer-specific
data in a separate repository.

Until the broker-dealer’s internal applications are able to consume CDM business
events directly, the broker-dealer would have to transform CDM business events into
its internal data model (IDM) business events for its applications to consume. Such
IDM business events are typically in an existing proprietary format and are persisted
in an internal data store, so it should be noted that the CDM has synonyms to allow
firms to internally extend the CDM to map to their IDM and manage those mappings
over time until the firm transitions to using the CDM directly. The internal data store
may need to be reconciled with the broker-dealer’s ADS, so careful design is required
to minimize any increase in complexity arising from interim adoption scenarios.

We now consider some specific options that may be made available to broker-dealers
regarding their degree of integration with the central ADS. Figure 1 illustrates
four key options that can be viewed as different adoption scenarios along a long-term
journey of increasing integration, ie, the first three scenarios are interim states and
the fourth scenario is the end state.

\begin{itemize}
\item The {\em }“scenario of first interim state using centralized ADS” shows a broker-dealer
that has minimal integration with the central ADS. It receives CDM
business events and transforms them into IDM business events, which are
then carried through to its internal data store (this corresponds to an extract,
transform, load (ETL) pattern of extracting at the FMI ADS and then transforming
and loading at the broker-dealer). The IDM business events are
then consumed by the broker-dealer’s trade applications. The CDM business
events are also logged. Note that there is no local ADS and the trade
applications remain unchanged. Reconciliation may be required between the
broker-dealer’s internal data store and the CDM business events log.

\vspace{3mm}

\begin{figure}[h!]
  \begin{center}
  \includegraphics[height=17.5cm, width=16.5cm]{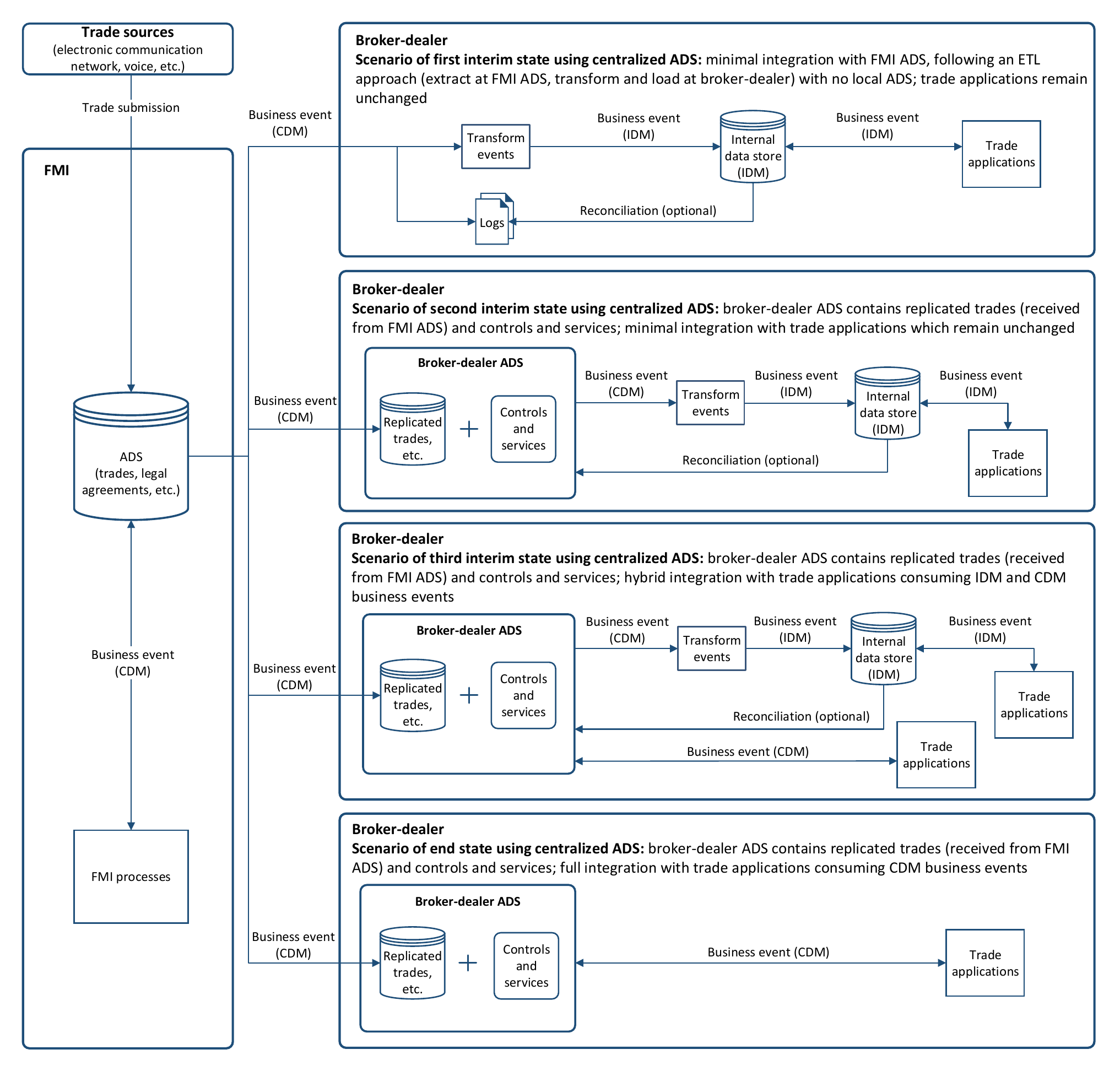}
  \vspace{5mm}
  
  \parbox{15 cm}{\caption{\footnotesize{ Centralized model showing an FMI maintaining a central ADS and generating CDM business events that are consumed by broker-dealers.
       \label{fig:centralised-model}}}}
  \end{center}
\end{figure}
  \pagebreak
                                                 
\item The {\em }“scenario of second interim state using centralized ADS” shows a broker-dealer
that receives CDM business events and maintains its own local ADS
including replicated trades from the central ADS. It retrieves CDM business
events from its local ADS and transforms them into IDM business events,
which are then carried through to its internal data store and consumed by its
trade applications, which remain unchanged. Reconciliation may be required
between the broker-dealer’s internal data store and its local ADS.

\item The {\em }“scenario of third interim state using centralized ADS” shows a hybrid
model in which some of the broker-dealer’s trade applications continue to consume
IDM business events from its internal data store (as per the “scenario
of second interim state using centralized ADS”) but other trade applications
consume CDM business events directly from its local ADS (which includes
replicated trades from the central ADS). This scenario illustrates the additional
flexibility (resulting in some additional complexity) that may be required while
a broker-dealer is transitioning from an IDM to the CDM.

\item The {\em }“scenario of end state using centralized ADS” shows a broker-dealer that
has fully integrated with its local ADS (which includes replicated trades from
the central ADS) and fully adopted the CDM internally, with its trade applications
consuming CDM business events directly from that local ADS. This is
the least complex centralized scenario and can be considered an end state for
the centralized model.

\end{itemize}

%\begin{figure}[h!]
%\begin{center}
%\includegraphics[height=17.5cm, width=16.5cm]{Central(v4.3).pdf}
%\vspace{5mm}

%\parbox{15 cm}{\caption{\footnotesize{ Centralized model showing an FMI maintaining a central ADS and generating CDM business events that are consumed by broker-dealers.
%     \label{fig:centralised-model}}}}
%\end{center}
%\end{figure}
%\pagebreak

\vspace{-2mm}

\subsection{Decentralized Model}
\label{sec:decentralizedModel}
\noindent
The set of architectural options for a decentralized model is typically more complicated
than for a centralized model. For example, instead of the ADS being maintained
centrally by the FMI, the ADS could be maintained by a combination of both
the FMI and the broker-dealers, using both distributed data and decentralized processing.
Overall synchronization could potentially be facilitated via emerging technologies
such as distributed ledger technology (DLT)\cite{uk-gov-dlt-beyond-blockchain},where a system
of distributed ledger nodes acting together can provide a single source of truth
among multiple parties. The Depository Trust \& Clearing Corporation is one of the
post-trade market infrastructures that is exploring the potential to leverage DLT as
a pivotal piece of technology that may help bring about new efficiencies in clearing
and settlement \cite{dtcc-progress-report-otc, dtcc-project-ion}.

Migrating from a centralized model to a decentralized model would involve significant
infrastructure changes, but such a journey can also be viewed as a series
of incremental stepping-stones. Figure 2 illustrates four of the options that can be
viewed as different adoption scenarios along a long-term journey of increasing integration
with an ADS and with DLT, ie, the first three scenarios are interim states and
the fourth scenario is the end state.

\begin{itemize}
\item The {\em }“scenario of first interim state using decentralized ADS” shows a broker-dealer
that has not adopted DLT and thus relies on the FMI to host its DLT
node on its behalf. Similar to the “scenario of first interim state using centralized
ADS” in the previous subsection, the broker-dealer receives CDM
business events and transforms them into IDM trade events, which are then carried through to its internal data store (this corresponds to an ETL pattern
of extracting at the “broker-dealer’s DLT node” hosted at the FMI and
then transforming and loading at the broker-dealer). The IDM business events
are then consumed by the broker-dealer’s trade applications. The CDM business
events are also logged. Note that there is no local ADS and the trade
applications remain unchanged. Reconciliation may be required between the
broker-dealer’s internal data store and the CDM business events log.

\begin{figure}[h!]
  \begin{center}
  \includegraphics[height=17.5cm, width=16.5cm]{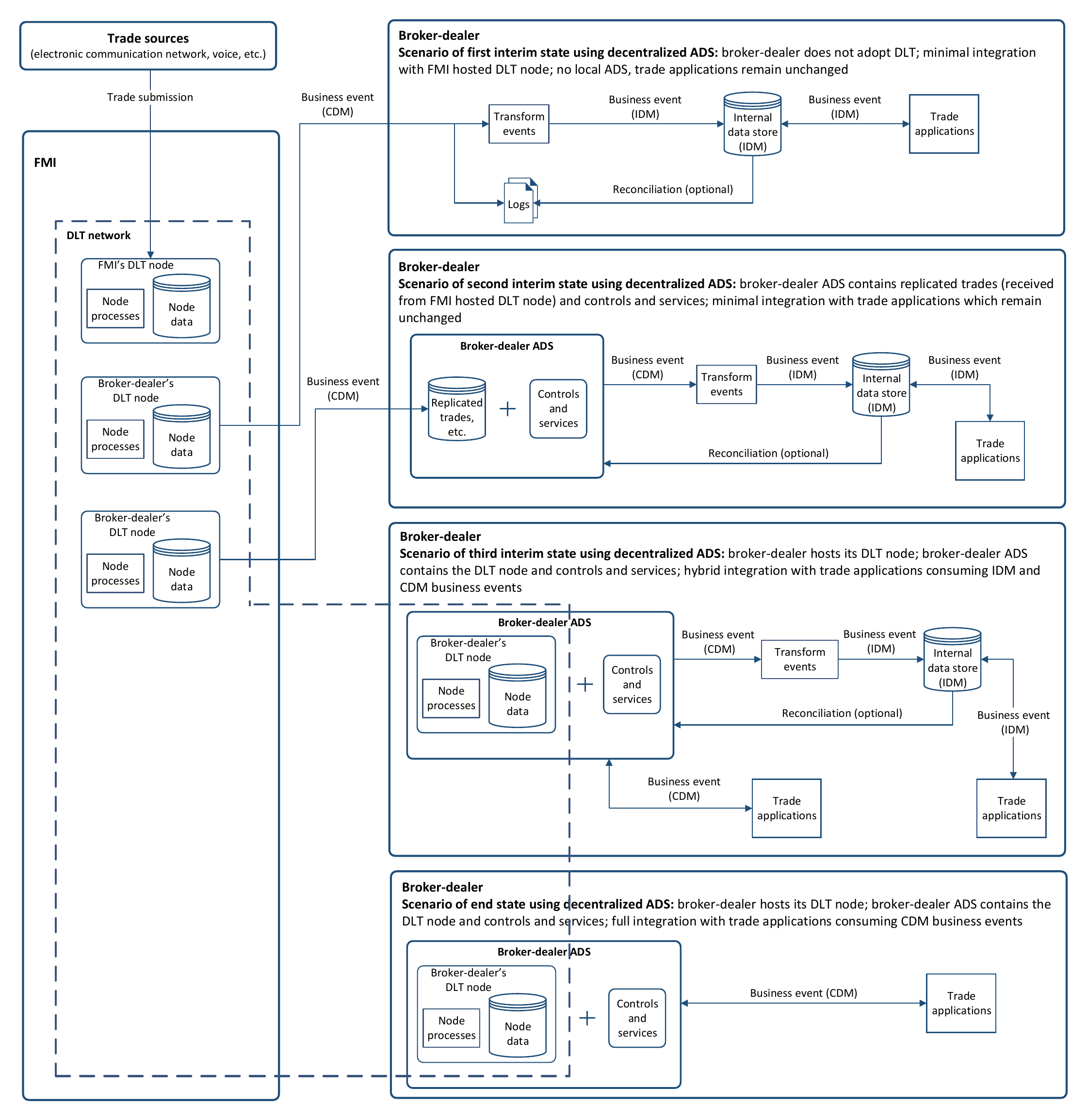}
  \vspace{5mm}
  
  \parbox{15 cm}{\caption{\footnotesize{ Decentralized model showing an FMI maintaining a distributed ADS with broker-dealers and processing CDM business events.
       \label{fig:decentralised-model}}}}
  \end{center}
  \end{figure}
  \pagebreak

\item The {\em }“scenario of second interim state using decentralized ADS” shows a
broker-dealer that receives CDM business events and maintains its own ADS
including replicated trades from the central ADS. Similar to the “scenario of
second interim state using centralized ADS” in the previous subsection, the
broker-dealer retrieves CDM business events from its local ADS and transforms
them into IDM business events, which are then carried through to
its internal data store and consumed by its trade applications, which remain
unchanged. Reconciliation may be required between the broker-dealer’s internal
data store and its local ADS. Similar to the “scenario of first interim state
using decentralized ADS”, the broker-dealer has not adopted DLT and thus
relies on the FMI to host the node on its behalf.

\item The {\em }“scenario of third interim state using decentralized ADS” shows a broker-dealer hosting its DLT node and therefore taking on greater responsibility
within the DLT network, with the FMI continuing to be an operator of the
DLT network including setting the rules. The local ADS contains the broker-dealer’s
DLT node. Similar to the “scenario of third interim state using centralized
ADS”, there is a hybrid model in which some of the broker-dealer’s trade
applications continue to consume IDM business events from its internal data
store but other trade applications consume CDM business events directly from
its local ADS (which includes trades in its DLT node). This scenario illustrates
the additional flexibility (resulting in some additional complexity) that may be
required while a broker-dealer is transitioning from an IDM to the CDM and
also the additional complexity that may be required while a broker-dealer is
transitioning to DLT.

\item The {\em }“scenario of end state using decentralized ADS” shows a broker-dealer
hosting its DLT node and having full integration with its local ADS (which
includes trades in its DLT node). Similar to the “scenario of end state using
centralized ADS”, the CDM has been fully adopted internally, with the trade
applications consuming CDM business events directly from the local ADS.
This is the least complex decentralized scenario and can be considered an end
state for the decentralized model.
\end{itemize}

%\begin{figure}[h!]
%\begin{center}
%\includegraphics[height=17.5cm, width=16.5cm]{Figure 2 - Decentralised model from Industry Adoption Scenarios for Authoritative Data Stores using the ISDA Common Domain Model.pdf}
%\vspace{5mm}

%\parbox{15 cm}{\caption{\footnotesize{ Decentralized model showing an FMI maintaining a distributed ADS with broker-dealers and processing CDM business events.
%     \label{fig:decentralised-model}}}}
%\end{center}
%\end{figure}
%\pagebreak

\section{Summary and Further Work}
\label{sec:summaryAndFurtherWork}	

In this paper, we highlighted an opportunity to significantly increase efficiency and
reduce costs in the post-trade industry. We discussed using the ISDA CDM standard
to address the industry problems of inconsistent processes and inconsistent data.
We also discussed using authoritative data stores to address the industry problem
of duplicated data. We then explored various adoption scenarios, considering both
traditional centralized models and potential decentralized models.

For further work, Barclays is developing prototypes of the end states for both the
centralized model and the decentralized model. We are reporting publicly on the
findings, including the prototype of the centralized model \cite{simulation-paper} and an
upcoming technical comparison between the two models. There is also opportunity
to explore the impact of regulatory requirements and liability concerns on a broker-dealer’s
ability or willingness to adopt the proposed changes. There are many design
choices when architecting industry ADSs using the CDM and we look forward to
continuing industry collaboration to develop good patterns and frameworks.

\vspace{7.5mm}
\noindent \textbf{ACKNOWLEDGEMENTS:} 
\noindent
We thank Ian Sloyan (ISDA), Rajagopalan Siddharthan (Barclays) and Vikram
Bakshi (Barclays) for their helpful feedback.
\vspace{5mm}

% Return to 'normal' rules on hyphenation for bibliography
\pretolerance=-1
\tolerance=-1
\emergencystretch=0pt

\bibliography{CDMPaper}
\bibliographystyle{plain}

\end{document}